\documentclass[twocol]{ametsocV6.1}

\usepackage{array, makecell}

\title{A simple model for predicting tropical cyclone minimum central pressure from intensity and size \\
{\color{red}NOT PUBLISHED. In peer review} }

\authors{Daniel R. Chavas\aff{a}\correspondingauthor{Daniel R. Chavas, drchavas@gmail.com}
John A. Knaff,\aff{b} 
Philip J. Klotzbach,\aff{c} 
}

\affiliation{\aff{a}{Purdue University, Department of Earth, Atmospheric, and Planetary Sciences, West Lafayette, IN}\\
\aff{b}{NOAA Center for Satellite Applications and Research, Fort Collins, CO}\\
\aff{c}{Colorado State University, Department of Atmospheric Science, Fort Collins, CO}
}

\abstract{Minimum central pressure ($P_{min}$) is an integrated measure of the tropical cyclone wind field and is known to be a useful indicator of storm damage potential. A simple model that predicts $P_{min}$ from routinely-estimated quantities, including storm size, would be of great value. Here we present a simple linear empirical model for predicting $P_{min}$ from maximum wind speed, the radius of 34-knot winds ($R_{34kt}$), storm-center latitude, and the environmental pressure. An empirical model for the pressure deficit is first developed that takes as predictors specific combinations of these quantities that are derived directly from theory, based on gradient wind balance and a modified-Rankine-type wind profile known to capture storm structure inside of $R_{34kt}$. Model coefficients are estimated using data from the southwestern North Atlantic and eastern North Pacific from 2004--2022 using aircraft-based estimates of $P_{min}$, Extended Best Track data, and estimates of environmental pressure from Global Forecast System (GFS) analyses. The model has near-zero conditional bias even for low $P_{min}$, explaining 94.4\% of the variance. Performance is superior to a variety of other model formulations, including a standard wind-pressure model that does not account for storm size or latitude (89.4\% variance explained). Model performance is also strong when applied to high-latitude data and data near coastlines. Finally, the model is shown to perform comparably well in an operations-like setting based solely on routinely-estimated variables, including the pressure of the outermost closed isobar. Case study applications to five impactful historical storms are discussed. Overall, the model offers a simple and fast prediction for $P_{min}$ for practical use in operations and research.}

\begin{document}

\maketitle


\statement
Sea level pressure is lowest at the center of a hurricane and is routinely estimated in operational forecasting along with the maximum wind speed. While the latter is currently used to define hurricane intensity, the minimum pressure is also a viable measure of storm intensity that is known to better represent damage risk. A simple empirical model that predicts the minimum pressure from maximum wind speed and size doesn't currently exist. This work develops such a model by using wind fied physics to determine the important parameters and then uses a simple statistical model to make the final prediction. This model is quick and easy to use in weather forecasting and risk assessment applications.

%
%
%
%
%
%

%




\section{Introduction}




At 21 UTC on 22 September 2005, the National Hurricane Center forecast discussion for Hurricane Rita stated "The minimum central pressure has remained around 913 mb... which is a very low pressure to have only 125 knots." This statement reflects a relatively infrequent but important issue when the minimum central pressure in a tropical cyclone differs strongly from what would be ``expected'' from its maximum wind speed alone based on historical experience. This example emphasizes the confusion that arises when our two common measures of tropical cyclone intensity -- the maximum wind speed $V_{max}$, which is the official measure of the Saffir-Simpson Hurricane Wind Scale, and the minimum central pressure $P_{min}$ -- depart from one another. This confusion begins with the attempt to characterize a storm's strength scientifically but then extends to the translation of this characterization into potential implications for the general public.

Indeed, it has long been standard to convert between the tropical cyclone maximum wind speed and $P_{min}$ using a simple empirical ``wind-pressure'' relation \citep{Dvorak_1975,Dvorak_1984,Atkinson_Holliday_1977,Koba_etal_1990,Knaff_Zehr_2007,Courtney_Knaff_2009}. This one-to-one relation assumes that $P_{min}$ depends predominantly on $V_{max}$. \cite{Knaff_Zehr_2007} demonstrated that minimum pressure also depends secondarily on storm size and latitude, motivated by gradient wind balance. This latter result was explained physically by \cite{Chavas_Reed_Knaff_2017} by combining gradient wind balance with a theoretical wind structure model. \cite{Chavas_Reed_Knaff_2017} demonstrated that a simple linear model that depends on $V_{max}$ and the product of outer size and the Coriolis parameter successfully predicted the central pressure deficit in simulations and observations. However, that work employed as its size metric the radius of $8 \: ms^{-1}$, which is not routinely estimated operationally nor reanalyzed for retention in a long-term archive. As a result, the utility of this model has been limited. A predictive model aligned with \citet{Chavas_Reed_Knaff_2017} that takes as input parameters that are routinely estimated in operations would be much more useful.


The potential utility of a precise and easy-to-use model for $P_{min}$ has grown as recent work has directly connected $P_{min}$ to risk. $P_{min}$ is a remarkably good predictor of the normalized economic damage that has been wrought by landfalling hurricanes in the continental United States \citep{Bakkensen_Mendelsohn_2016,Klotzbach_etal_2020,Klotzbach_etal_2022}. \cite{Klotzbach_etal_2020} demonstrated that $P_{min}$ is a substantially better predictor than $V_{max}$, particularly for hurricanes making landfall from Georgia to Maine where weaker but larger storms are more common. \cite{Klotzbach_etal_2022} demonstrated that $P_{min}$ is also at least as good of a predictor as integrated measures of the near-surface wind field, including both integrated kinetic energy and integrated power dissipation, that inherently require data from the entire wind field to estimate. The explanation lies in the fact that $P_{min}$ is itself an integrated measure of the wind field that accounts for both maximum wind speed and storm size \citep{Chavas_Reed_Knaff_2017}. The total wind field drives the wind, storm surge, and rainfall hazards that ultimately cause damage and loss of life \citep{Irish_Resio_2010,Zhai_Jiang_2014}. Hence, $P_{min}$ appears to be an especially well-suited measure of the damage potential of a storm, and it carries ancillary practical benefits. 

First, $P_{min}$ is relatively easy to estimate, as it requires relatively few observations within a small area near the storm center and, moreover, it varies relatively smoothly in space and time since it is by definition an integrated quantity (either in radius via gradient wind balance or in height via hydrostatic balance). In contrast, $V_{max}$ is a local estimate at a single point of a quantity that is inherently noisy and hence is notoriously difficult to estimate from sparse observations \citep{Uhlhorn_Nolan_2012}. Second, $P_{min}$ is already routinely estimated operationally; indeed, it was used in conjunction with maximum wind speed as part of the Saffir-Simpson scale prior to 2009 \citep{Schott_etal_2012}.

The practical benefits of a model for $P_{min}$ extend beyond operations to climate and risk modeling. Climate models are better able to reproduce the historical distribution of minimum pressure than maximum wind speed \citep{Knutson_etal_2015}, suggesting that the former is a more stable and suitable metric for model evaluation and intercomparison \citep{Zarzycki_Ullrich_Reed_2021}. For risk modeling, the pressure field is included as input in storm surge models \citep{Gori_etal_2023}. More generally, a model that can relate $P_{min}$, $V_{max}$, and size to one another should enable the use of all available data to more precisely constrain the properties of historical tropical cyclones and their relationships to hazards and potential impacts.

Here our objective is to create a simple model to predict $P_{min}$ that can be easily used in operations and practical applications. The spirit of this effort to make theory directly useful for the community follows from a similar effort for predicting $R_{max}$ presented in \cite{Chavas_Knaff_2022} and \cite{Avenas_etal_2023}. Section \ref{sec:methods} describes the datasets used in our analysis. Section \ref{sec:model} develops the empirical model and its physical basis. Section \ref{sec:results} estimates model parameters from data and applies the model in a few notable contexts, including an operational setting using only routinely-estimated parameters. Section \ref{sec:conclusions} provides a brief summary and discussion.

\section{Data and Methods}\label{sec:methods}

Our dataset combines flight-based data for $P_{min}$, final best track data for storm central latitude, maximum wind speed $V_{max}$, quadrant-maximum radius of 34-kt wind ($R_{34kt}$), and estimates of environmental pressure $P_{env}$ from analysis and best track data, We analyze storms in both the North Atlantic and eastern North Pacific basins for the period 2004-2022, where 2004 is the first year in which post-season best tracking of $R_{34kt}$ was performed. All information is contained in the databases of the Automated Tropical Cyclone Forecast system  \citep [ATCF;][]{Sampson_Schrader_2000}. These data are identical to that available in the Extended Best Track \citep[EBTRK,][]{Demuth_DeMaria_Knaff_2006}.

Direct observational estimates of $P_{min}$ are critical to ensure that these data are true independent observations and are not inferred from $V_{max}$ via an existing wind-pressure relationship (which is difficult to determine in retrospect), as this would undercut the purpose of our work. Hence, we take $P_{min}$ estimates based on aircraft and recorded in the ATCF databases. We then linearly interpolate best track data to the times of flight-based estimates of $P_{min}$. We calculate the mean radius of 34-kt wind at a given time, hereafter $\overline{R}_{34kt}$, by averaging all non-zero quadrant values and then multiplying the result by 0.85 to account for the fact that the quadrant values are defined as the maximum extent in that quadrant and are not the mean. A similar approach was employed by \cite{DeMaria_etal_2009} and \cite{Chavas_Knaff_2022}.

A combination of the NCEP Climate Forecast System  \citep[CFS;][]{sahaEA2014} and Global Forecast System (GFS) analyses \citep{GFS2021} were used to estimate $P_{env}$.  CFS analyses were used in 2004 and 2005, and GFS analyses were used thereafter.  We use the profile of tangential wind in the analysis, specifically the radius of 8 $m^{-1}$ ($r8$), to inform us what radii represents the outer edge of the storm and to calculate $P_{env}$. $P_{env}$ was calculated at 900, 1200, 1500, 1800, and 2100 km for ranges of $r8$ of 0 to 600 $km$, 600 to 900 $km$, 900 to 1200 $km$, 1200 to 1500 $km$ and greater than 1500 km, respectively.  In the final section, we also estimate the environmental pressure for operational relevance using the pressure of the outermost closed isobar ($P_{oci}$) extracted from the ATCF databases or EBTRK dataset.   

Following \cite{Chavas_Knaff_2022}, we filter the data to focus on cases over the open ocean within the tropical western Atlantic ocean where aircraft reconnaissance is routine. Hence, we restrict our data to cases west of $50 \; ^oW$ and south of $30 \; ^oN$ and with $\overline{R}_{34kt}$ smaller than the distance from the storm center to the coastline. We only include cases with at least three non-zero quadrant values of $R_{34kt}$ to ensure a reasonable estimate of the azimuthal-mean value, and where $V_{max}$ is at least $20 \; ms^{-1}$ to avoid weak tropical storms and tropical depressions. A map of the final dataset is displayed in Fig. \ref{fig:mapmodelalldata}a.

\section{Model for pressure deficit $\Delta P$}\label{sec:model}

The minimum pressure is given by:
\begin{equation}\label{eq:PmindP}
    P_{min} = P_{env} + \Delta P
\end{equation}
where $P_{min}$ is the minimum central pressure, $P_{env}$ is the environmental pressure, and $\Delta P$ is the pressure deficit at the storm center (defined as negative for low pressure). Here we develop a model specifically for $\Delta P$, as this is what is tied physically to the wind field. We also test $P_{min}$ alone below and demonstrate that knowledge of $P_{env}$ does add additional value. Practical use of the model is tackled in Section 4c below, where we combine our model prediction for $\Delta P$ with an estimate for $P_{env}$ to predict $P_{min}$ itself.

We develop a model for $\Delta P$ that is derived from gradient wind balance applied to a two-region, modified-Rankine vortex model of the axisymmetric tropical cyclone wind field. The full derivation is presented in Appendix A. Here we focus on the core outcome from the theory. $\Delta P$ depends linearly on three physical predictors: $V_{max}^2$, $\frac{1}{2}f\overline{R}_{34kt}$, and $\frac{\frac{1}{2}f\overline{R}_{34kt}}{V_{max}}$. $V_{max}$ is the maximum wind speed, $\overline{R}_{34kt}$ is the radius of 34-kt winds (i.e., the outermost radius routinely estimated operationally), and $f$ is the Coriolis parameter at the storm center latitude. Because the theory has various approximations built into in it, we do not use the final theoretical result (Eq \eqref{eq:dp_analytic_numerical}) directly, but rather we use theory simply to identify the predictors. We then model $\Delta P$ using a simple multiple linear regression (MLR) model using those predictors:
\tiny
\begin{equation}\label{eq:mlr_basic}
    \Delta P_{hPa} = \beta_0 + \beta_{Vmax^2}\left(V_{max}^2\right) + \beta_{fR}\left(\frac{1}{2}f\overline{R}_{34kt}\right) + \beta_{fRdV}\left(\frac{\frac{1}{2}f\overline{R}_{34kt}}{V_{max}}\right)
\end{equation}
\normalsize
This approach produces an optimal model that exploits the benefits of both theory and observational data. A similar approach was taken to predict the radius of maximum wind in \citet{Chavas_Knaff_2022} and \citet{Avenas_etal_2023}.

Briefly, we note the contrast with \cite{Chavas_Reed_Knaff_2017}, which found a simpler multiple linear regression relationship for $\Delta P$ on $V_{max}$ and $\frac{1}{2}fR$, where $R$ was taken as the radius of $8 \: ms^{-1}$. These two parameters were taken directly from the wind structure theory of \cite{Chavas_Lin_2016} applied to gradient wind balance, and indeed they show up here too. However, since the structural theory of \cite{Chavas_Lin_2016} has no analytic solution, \cite{Chavas_Reed_Knaff_2017} could not derive an analytic solution for $P_{min}$, and hence could not define exactly how $P_{min}$ should depend on those parameters (their Eq. 12). Instead, they merely found that a pure linear dependence worked quite well. The new model presented here (Eq. \eqref{eq:mlr_basic}) now gives that analytic dependence. This dependence corroborates the results of \cite{Chavas_Reed_Knaff_2017}. We find that a pure linear dependence does work well (same as the first two terms in Eq. \eqref{eq:mlr_basic}) but with a slight deviation associated with the third term in Eq. \eqref{eq:mlr_basic} that depends on the ratio of the two parameters. Note that while it might seem desirable to take the reciprocal of the third term, this will make the dependence on the term nonlinear, and indeed doing so results in significantly degraded performance. For physical interpretation of the three terms, see Appendix A. 

To estimate model coefficients (Eq. \eqref{eq:mlr_basic}), we first bin the dataset into increments of $10 \: ms^{-1}$ for $V_{max}$, $50 \; km$ for $\overline{R}_{34kt}$, and $10^{-5} \: s^{-1}$ for f (i.e. approximately $4^o$ latitude), resulting in a single datapoint per joint $(V_{max}, \overline{R}_{34kt}, f)$ bin. Binning minimizes the effects of variations in sample size across the phase space of these variables when estimating model parameters, and it also helps reduce noise within a given bin. No minimum sample size is imposed within each bin, so all bins with at least one valid datapoint are retained. Multiple linear regression coefficients are calculated using the fitlm function in MATLAB. We use the entire dataset in order to produce the best estimate of the model coefficients from all available data. We estimate the 95\% confidence interval of each model coefficient as the 2.5$^{th}$ and 97.5$^{th}$ percentile from a 1000-member bootstrap with resampling of the raw dataset, each fit following the same procedure as above. We explicitly account for the number of degrees of freedom in model performance metrics by calculating the adjusted r-squared value reduced from the nominal value based on the number of degrees of freedom relative to the sample size, $r^2_{adj} = 1-[(n-1)/(n-p)](SSE/SST)$ where $n$ is the sample size, $p$ is the number of coefficients, $SSE$ is the sum of squared error, and $SST$ is the sum of squared total. The adjustment factor $(n-1)/(n-p)$ is very small ($\approx 1.0015$, i.e. a $0.15\%$ reduction from nominal) since the number of coefficients $p=4$ is much smaller than our sample size of approximately 2000, indicating that model overfitting is not a concern. This approach is analogous to removing an arbitrary subset of years for out-of-sample validation, and it carries the added benefit of allowing us to use the entire dataset to yield the best possible coefficient estimates.  To evaluate model performance, we apply the model described above to the full, raw dataset and quantify the statistics of the observed vs. predicted $\Delta P$ or $P_{min}$.

We first present the final model and its performance. We then demonstrate the utility of each term in the model, including a comparison with a standard ``wind-pressure'' model in which $V_{max}$ is the only predictor. We then apply the model to a few specific out-of-sample applications that are of practical interest. Finally, we show how to use the model in an operational setting using only routinely-estimated data and examine model performance for a few case studies of impactful storms in the recent historical record.

\section{Results}\label{sec:results}

\subsection{Model results}

\begin{figure*}[t]
\centering
 \noindent\includegraphics[width=0.5\textwidth]{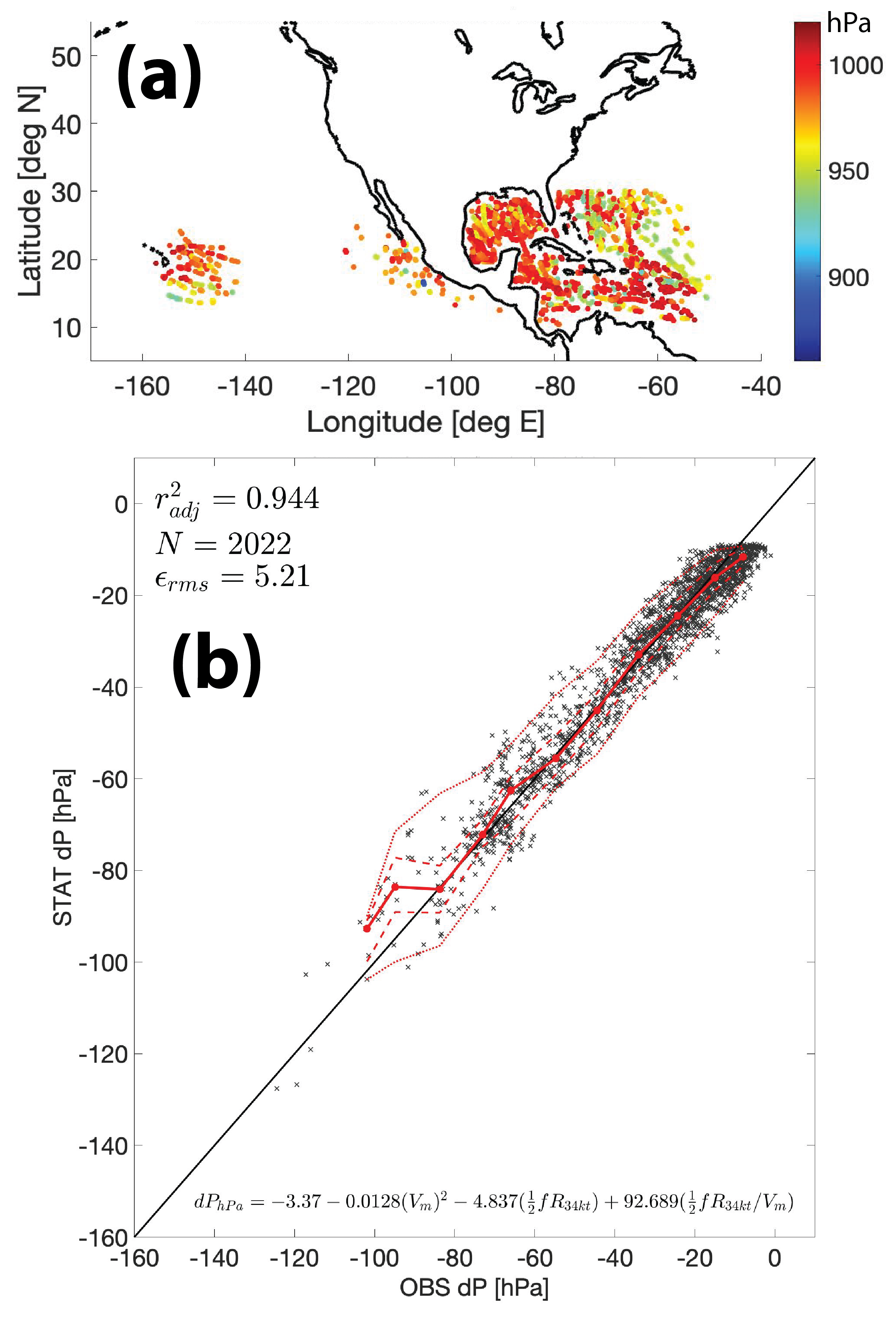}\\
 \caption{(a) Map of the aircraft-based historical $P_{min}$ dataset used in this study; the color denotes the magnitude of $P_{min}$ [hPa]. (b) Model prediction for $\Delta P$ (Eq. \eqref{eq:mlr_final}; y-axis) vs. observed $\Delta P$ (x-axis).}\label{fig:mapmodelalldata}
\end{figure*}

Our final model is:
\tiny
\begin{equation}\label{eq:mlr_final}
    \Delta P_{hPa} = -3.37 - 0.0128\left(V_{max}^2\right) - 4.837\left(\frac{1}{2}f\overline{R}_{34kt}\right) + 92.689\left(\frac{\frac{1}{2}f\overline{R}_{34kt}}{V_{max}}\right)
\end{equation}
\normalsize
As described in Section \ref{sec:methods} above, the coefficient values are estimated by fitting the model to the dataset binned in fixed increments of $V_{max}$ $\overline{R}_{34kt}$, and $f$ to minimize the effect of sample size variability. The bootstrapped 95\% confidence interval range on each coefficient is (-4.44, -2.34) for $\beta_0$, (-0.0133, -0.0123) for $\beta_{Vmax^2}$, (-5.356, -4.537) for $\beta_{fR}$, and (84.045, 111.433) for $\beta_{fRdV}$.

Model performance is shown in Figure \ref{fig:mapmodelalldata}b, which displays predicted (Eq. \eqref{eq:mlr_final}) vs. observed $\Delta P$ for the full raw dataset shown in Figure \ref{fig:mapmodelalldata}a. Eq. \ref{eq:mlr_final} explains 94.4\% of the variance in $\Delta P_{hPa}$, with an RMS error of 5.21 hPa (Table \ref{tab:modelmods}). The model is nearly unbiased across the full range of $\Delta P_{hPa}$, as evident by the solid red line (binned median) in Figure \ref{fig:mapmodelalldata}b closely following the black one-to-one line. This unbiased behavior extends to the most extreme data points with the largest pressure deficits (lower-left region of figure).

The sign of the dependence on each parameter matches the theory. First and foremost, the model predicts larger pressure deficit ($\Delta P_{hPa}$) at higher intensity ($V_{max}^2$; positive dependence), and at larger size or higher latitude ($\frac{1}{2}f\overline{R}_{34kt}$; positive dependence), as has been shown in past work \citep{Knaff_Zehr_2007,Chavas_Reed_Knaff_2017}. The model also predicts a smaller pressure deficit for larger values of the ratio predictor ($\frac{\frac{1}{2}f\overline{R}_{34kt}}{V_{max}}$; negative dependence). This predictor is new so does not have a prior precedent. The third term is more complex, but given the positive coefficient, it yields a more intense storm (more negative pressure deficit) at higher $V_{max}$ or at smaller size or higher latitude (these both would make the term less positive) -- hence this term slightly enhances the dependence on $V_{max}$ (first term) but slightly reduces the dependence on $\frac{1}{2}f\overline{R}_{34kt}$ (second term). We show below that this new predictor has the smallest effect of the three predictors but still adds value to the model's predictive power.

\subsection{Tests of model formulation}

\begin{table*}[t]
    \centering
    \begin{tabular}{|c|c|c|c|} \hline 
         \textit{Model}&  \makecell{\textit{Expl. var.} \\ ($100 \times r^2_{adj}$)}& \makecell{\textit{Error} \\ ($\epsilon_{RMS}$ [hPa])} & \textit{MLR model coefficients} \\ \hline 
         \textbf{Final model (Eq. \eqref{eq:mlr_final})}&  \textbf{94.4}& \textbf{5.21}& \makecell{$(\beta_0, \beta_{Vmax^2},\beta_{fR},\beta_{fRdV}) = $ \\ \\ $(-3.37,-0.0128,-4.837,+92.689)$} \\ \hline 
         $V_{max}^2$ and $\frac{1}{2}f\overline{R}_{34kt}$ only&  92.9& 5.87& \makecell{$(\beta_0, \beta_{Vmax^2},\beta_{fR}) = $ \\ \\ $(0.74,-0.0152,-2.221)$} \\ \hline 
         $V_{max}^2$ only&  86.3& 8.18& \makecell{$(\beta_0, \beta_{Vmax^2},\beta_{fR},\beta_{fRdV}) = $ \\ \\ $(-3.37,-0.0128,-4.837,+92.689)$} \\ \hline 
         $V_{max}$ only&  88.7& 7.41& \makecell{$(\beta_0, \beta_{Vmax^2}) = $ \\ \\ $(-10.07,-0.0153)$} \\ \hline 
        Predict $P_{min}$ instead of $\Delta P$& 93.5&5.64& \makecell{$(\beta_0, \beta_{Vmax^2},\beta_{fR},\beta_{fRdV}) = $ \\ \\ $(1009.29,-0.0131,-4.390,+91.316)$} \\ \hline 
        \makecell{Predict $P_{min}$ = $\mathscr{F}(V_{max})$ \\ (classic wind-pressure)}& 89.4&7.17& \makecell{$(\beta_0, \beta_{Vmax}) = $ \\ \\ $(1038.19,-1.55)$} \\ \hline 
        Include 30-50N data& 93.7&5.38& \makecell{$(\beta_0, \beta_{Vmax^2},\beta_{fR},\beta_{fRdV}) = $ \\ \\ $(-3.16,-0.0129,-4.794,+91.992)$} \\ \hline 
        Translation LC12& 94.2&5.24& \makecell{$(\beta_0, \beta_{Vmax^2},\beta_{fR},\beta_{fRdV}) = $ \\ \\ $(-6.60,-0.0127,-5.506,+109.013)$} \\ \hline 
        Translation KZ07& 94.1&5.26& \makecell{$(\beta_0, \beta_{Vmax^2},\beta_{fR},\beta_{fRdV}) = $ \\ \\ $(-6.52,-0.0128,-5.457,+104.956)$} \\ \hline 

    \end{tabular}
    \caption{Model performance for our final multiple linear regression model (top line) and alternative versions to test the effect of modifications to the model formulation. See text for details.}
    \label{tab:modelmods}
\end{table*}


We next evaluate the effect of each predictor as well as a few choices made in the formulation of our final model. Changes in model performance for modifications to our model are shown in Table \ref{tab:modelmods}. In each case, the alternative model was fit and tested in an identical fashion as was done for our full model above.

First, we remove the least valuable predictor one at a time and present the model results for direct comparison to the full model. Removing the third predictor (ratio) increases the unexplained variance from 5.6\% to 7.1\%. Further removing the second predictor (size and latitude), i.e. a linear regression on $V_{max}^2$ alone, further increases the unexplained variance to 13.7\%. Finally, replacing $V_{max}^2$ with $V_{max}$ partially reduces the unexplained variance back to 11.3\%.

We also tested the same model but using $P_{min}$ rather than $\Delta P$ as the predictand (i.e., ignoring variations in $P_{env}$), which results in a slight reduction of performance with an increase in unexplained variance from 5.6\% to 6.5\%. This effect is relatively small but does indicate that knowledge of variations in the environmental pressure can help modestly improve the prediction of $P_{min}$ for a given storm.

Lastly, we combine the above two tests to examine a standard wind-pressure relationship, i.e. predicting $P_{min}$ from $V_{max}$ alone. This leaves 10.6\% of the variance unexplained. This model actually performs better than the two models that predict $\Delta P$ from $V_{max}$ or $V_{max}^2$. This result gives insight into why the standard wind-pressure relationship has been so successful, as apparently $V_{max}$ may capture some of the residual variance associated with the other two terms and variations in environmental pressure. Overall, our full model offers a significant improvement in performance over a standard wind-pressure relationship by capturing nearly half of the residual variance in the latter (94.4\% vs. 89.4\% variance explained).

Note that we also tried a range of other model fits with $V_{max}$, $\overline{R}_{34kt}$ and $f$ arbitrarily employed in a variety of ways both linearly and non-linearly. However, none outperformed our theory-driven version presented above. This does not outright prove that the theory-derived set of parameters found here are the ``optimal'' parameters, but it does lend confidence that theory really can provide useful guidance on how the observed parameters are best combined for the purposes of predicting the pressure deficit.

Finally, we test two other choices in our model formulation. First, the above outcomes hold true for the model fit to the same dataset but including all cases up to 50$^o$N. Its performance (93.7\% variance explained) is slightly degraded relative to our model up to 30$^o$N, which is unsurprising given the greater complexity of cases at higher latitudes, a topic we return to below. Second, we chose not to reduce the Best Track $V_{max}$ to account for translation speed effects as has been done in the past. Models that incorporate the two existing methods for reducing Best Track $V_{max}$ to account for this effect both reduce performance. The first method, from \cite{Lin_Chavas_2012}, applies a simple linear reduction of $V_{max}$ by 0.55 times the translation speed; this method very slightly reduces the explained variance to 94.2\%. The second method, from \cite{Knaff_Zehr_2007}, applies a non-linear modification based on the translation speed; this method also  slightly reduces the explained variance to 94.1\%. This outcome does not indicate that translation effects do not exist, as its right-of-track enhancement of the local wind speeds is well known, but rather likely reflects that its magnitude may vary significantly in space and time. Its implicit inclusion in $V_{max}$ may also vary significantly (e.g. due to observational inhomogeneities or variability across storms). As a result, its effect may be real but noisy enough that simple attempts to remove it generically across all cases do not improve the model.

\subsection{Applications to special subsets}

\subsubsection{High latitude storms} 

\begin{figure*}[t]
\centering
 \noindent\includegraphics[width=0.5\textwidth]{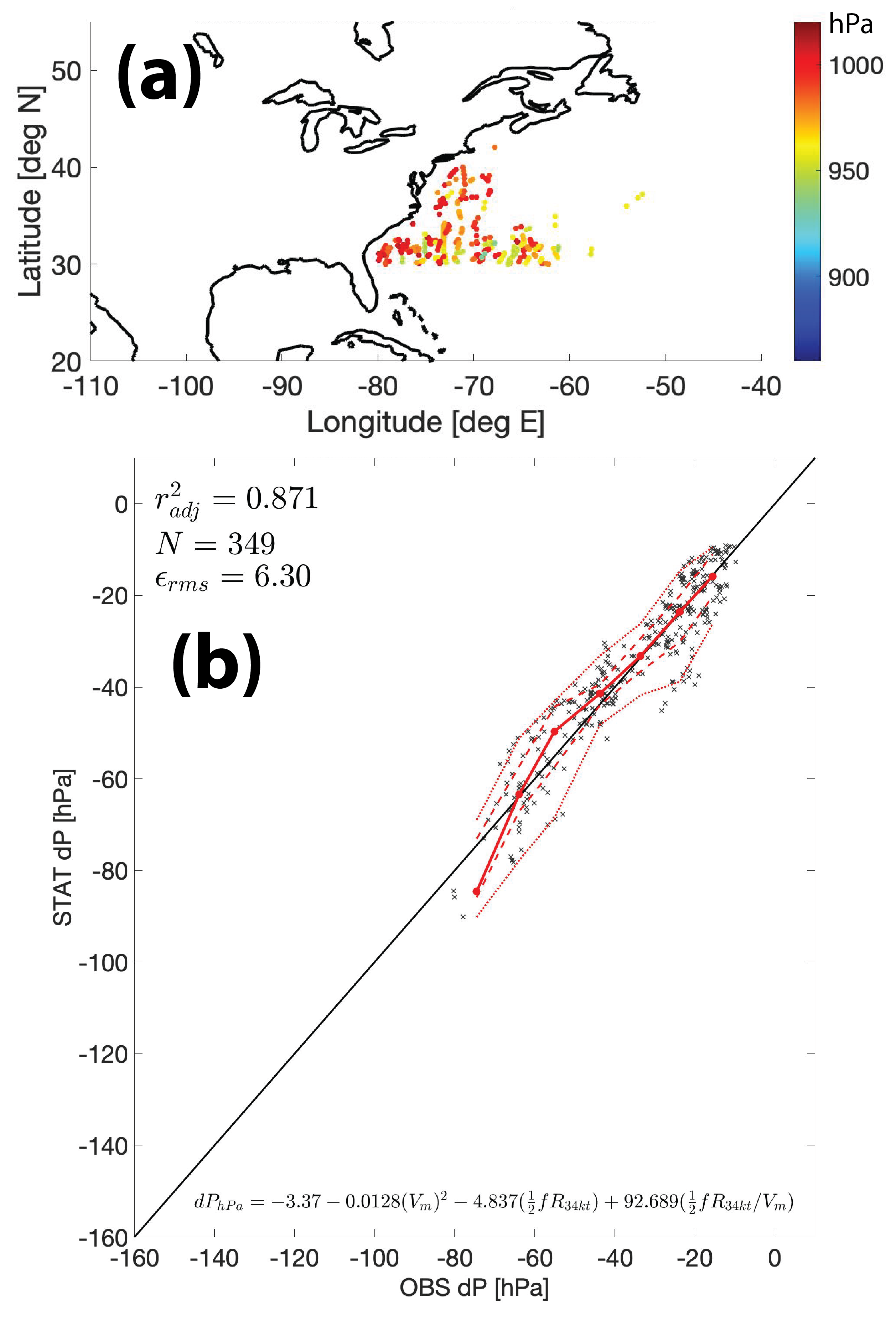}\\
 \caption{As in Figure \ref{fig:mapmodelalldata} but for $\Delta P$ predicted by our final model (Eq. \eqref{eq:mlr_final}) for data between 30$^o$N and 50$^o$N.}\label{fig:highlat}
\end{figure*}

We next apply our model to the data for high-latitude storms spanning 30-50$^o$N in our database (Figure \ref{fig:highlat}). At these higher latitudes, jet stream interactions and the onset of extratropical transition are much more likely, storms tend to expand, and North Atlantic storms often have significant interactions with land to the west. Hence, data in this region are associated with much larger uncertainty in both input parameter estimates $V_{max}$ and $\overline{R}_{34kt}$ and structural relationships among parameters. Nonetheless, the model performs reasonably well for this generally more complex subset of cases, explaining 87\% of the variance with an RMS error of 6.3 hPa. The model provides a nearly unbiased estimate of $\Delta P$ for values down to -60 hPa. For the strongest storms, though, there exists a systematic overestimation of $\Delta P$. Overall, while the performance is clearly worse than when applied to lower latitude storms as expected, the model appears to extrapolate well to higher latitudes. For further context, the high-latitude case of Hurricane Sandy is explored in the case studies section of this paper.

\subsubsection{Land Proximity Storms}

\begin{figure*}[t]
\centering
 \noindent\includegraphics[width=0.5\textwidth]{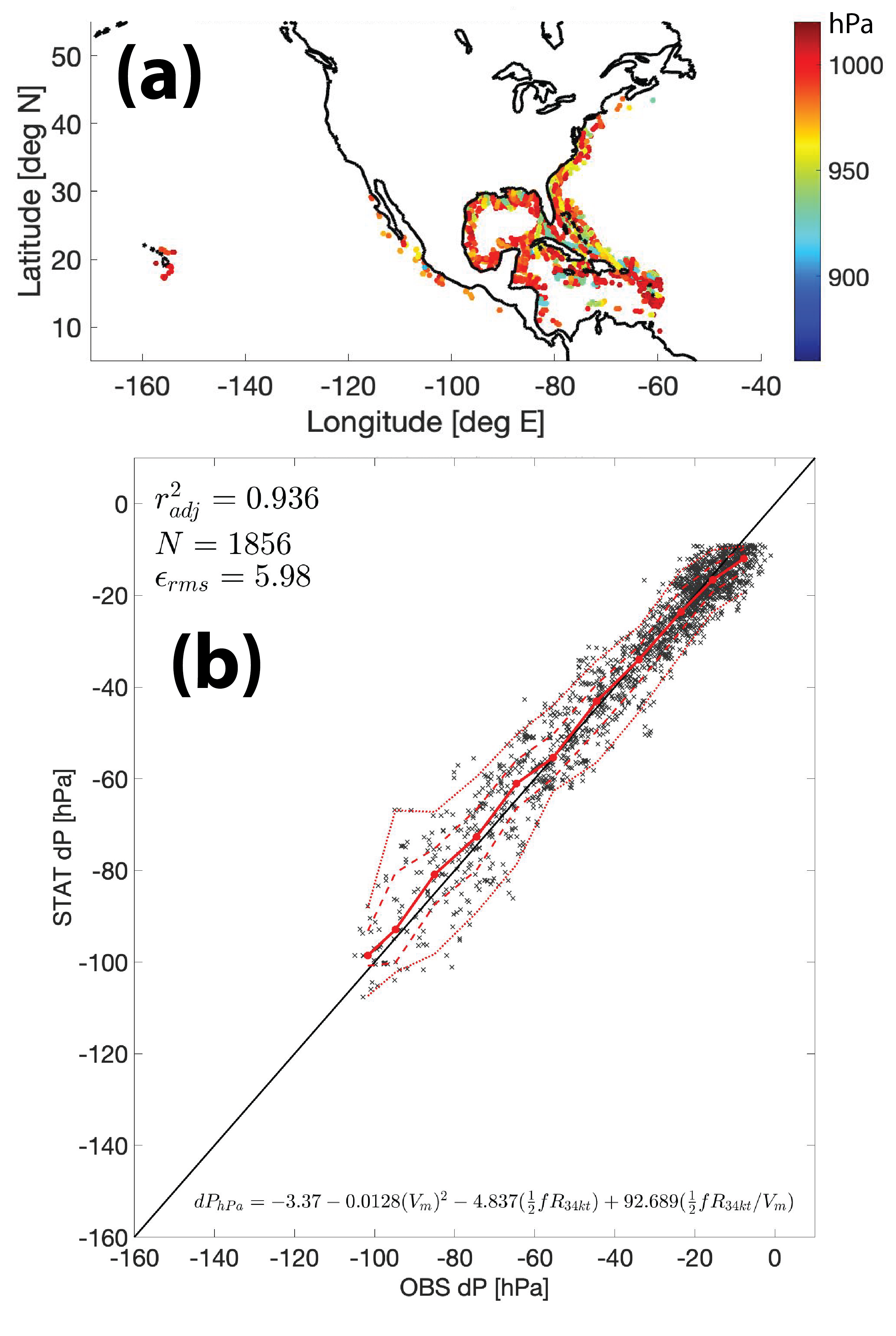}\\
 \caption{As in Figure \ref{fig:mapmodelalldata} but for $\Delta P$ predicted by our final model (Eq. \eqref{eq:mlr_final}) for data close to land, defined as within 200 km of a coastline.}\label{fig:closetoland}
\end{figure*}

Finally, we apply our model specifically to tropical cyclones that are relatively close to land and hence potentially of greater risk for coastal populations. Land introduces significant asymmetry in the surface wind field that would be expected to increase uncertainty, particularly in estimation of the azimuthal-mean $\overline{R}_{34kt}$. Results for our model application to all data within 200 km of a coastline up to 50$^o$N are shown in Figure \ref{fig:closetoland}. Model performance is again quite strong, explaining 93.6\% of the variance with an RMS error of 5.98 hPa and is nearly unbiased over the full range of pressure deficits. The error spread increases for stronger storms with pressure deficits $<$-60 hPa relative to weaker storms in this subset as well as comparable storms in our full dataset. Overall, though, the model also applies well for storms that are close to land.

\subsection{Practical application: Extended Best Track and historical case studies}

\begin{figure*}[t]
\centering
 \noindent\includegraphics[width=0.5\textwidth]{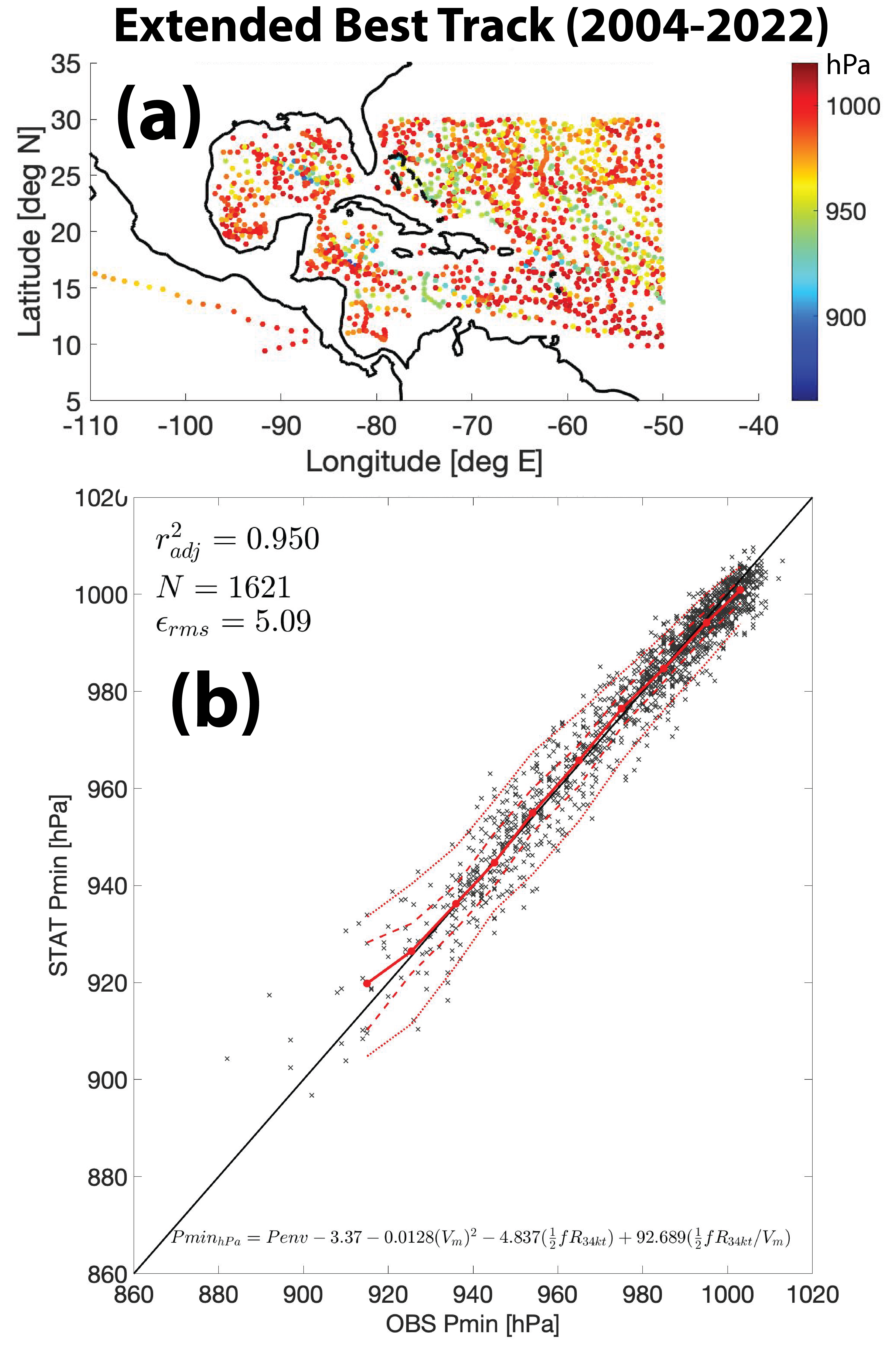}\\
 \caption{As in Figure \ref{fig:mapmodelalldata} but for the ``operational" model prediction, with $\Delta P$ predicted by our final model (Eq. \eqref{eq:mlr_final}) using only the Extended Best Track database for the period 2004-2022 in the North Atlantic for all data and $P_{env}$ estimated from $P_{oci}$ using Eq. \eqref{eq:Penv}.}\label{fig:ebtk}
\end{figure*}

We next demonstrate how the model can be put into direct practical use to predict $P_{min}$ using only operationally-available data. We use data exclusively from the EBTRK (2004-2022), which provides all parameters that are estimated in near-real time and are later refined in post-season best tracking. We fit the model to the North Atlantic only to give a more apples-to-apples comparison to our observation-based dataset that is weighted heavily towards the North Atlantic.

The lone parameter from our original analysis that was not already routinely estimated is the environmental pressure, $P_{env}$, which is needed to translate the model's prediction for $\Delta P$ back into $P_{min}$. In our prior analyses, we had estimated $P_{env}$ from the pressure field in global model reanalysis, which is more precise but not operationally available. Here we utilize the pressure of the outermost closed isobar ($P_{oci}$) and apply a simple constant offset
\begin{equation}\label{eq:Penv}
    P_{env} = P_{oci} + 2 \; hPa
\end{equation}
The value of 2 hPa was estimated directly from our model by finding the value that eliminates the mean bias in our prediction (an increase/decrease in $P_{env}$ simply acts to increase/decrease $P_{min}$ by the same amount). As shown below, this offset results in near zero conditional bias across all values of $P_{min}$, which indicates that using a simple constant for an offset is reasonable. Notably, this estimate of $P_{env}$ agrees with recommendations in \cite{Courtney_Knaff_2009}.


We predict $P_{min}$ by calculating $\Delta P$ using Eq. \eqref{eq:mlr_final} and $P_{env}$ using Eq. \eqref{eq:Penv}. The results are shown in Figure \ref{fig:ebtk}. The prediction compares quite well with the EBTRK data, explaining $95.0\%$ of the variance with an RMS error of 5.1 hPa. The performance is very similar to that found above using the observation-based database. This is to be expected, as EBTRK should mostly be very similar to our database, with the exception of EBTRK being at a six-hourly temporal resolution and for occasional times when aircraft reconnaissance were not available, which should be relatively infrequent by design given our focus on the western Atlantic.

These results indicate that one can make a good prediction for $P_{min}$ if given high-quality operational estimates of $V_{max}$, $\overline{R}_{34kt}$, storm center latitude, and  pressure of the outermost closed isobar. This may be of direct value when remote sensing-based wind speed measurements are available but measurements of $P_{min}$ are not.

\subsubsection{Historical case studies}

\begin{figure*}[t]
\centering
 \noindent\includegraphics[width=0.9\textwidth]{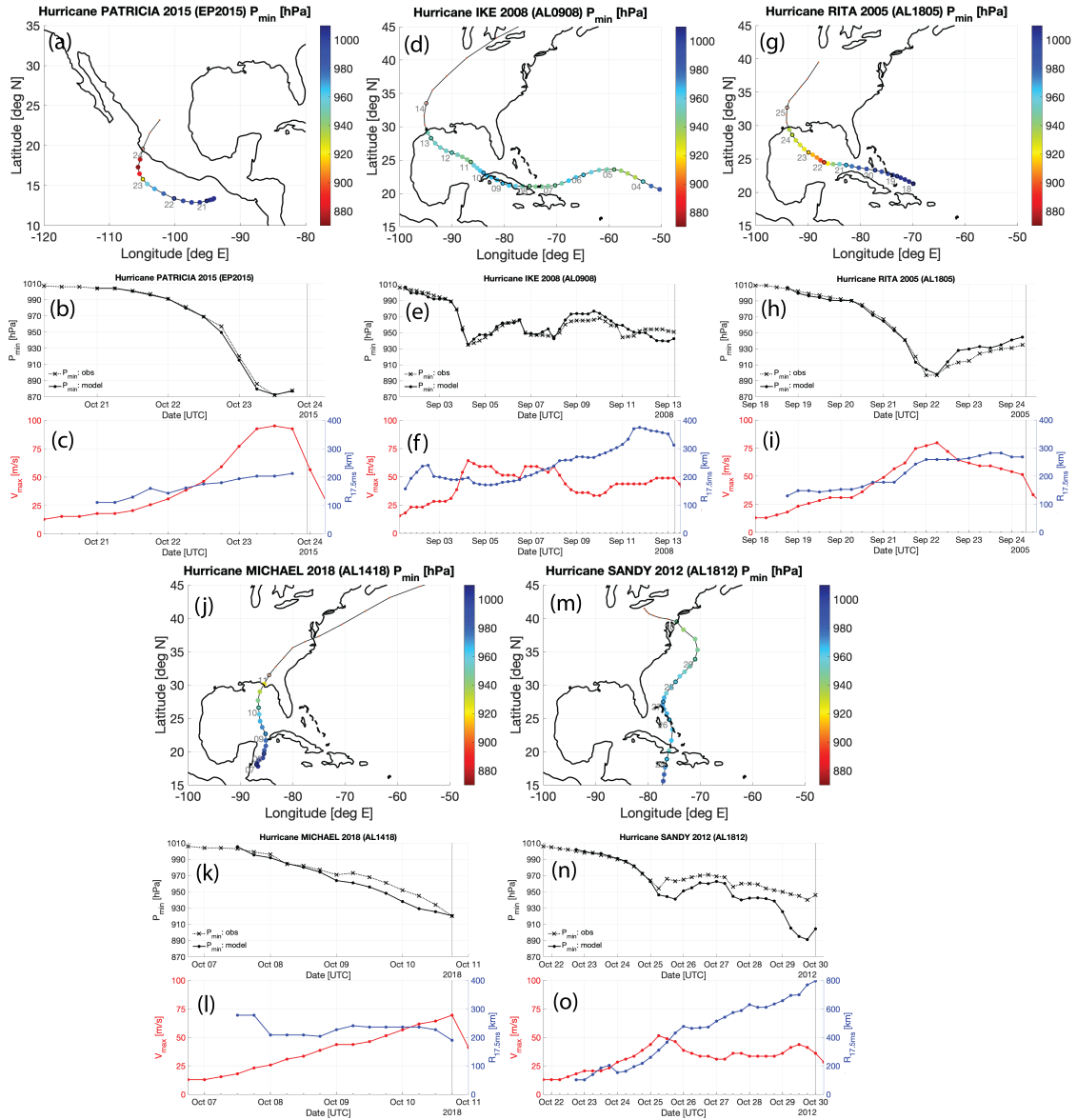}\\
 \caption{Model prediction for five case studies of recent impactful storms. Patricia (2015, EP): (a) map of track and $P_{min}$ (color); (b) Time series of Best Track $P_{min}$ vs. model prediction from Best Track operational inputs only; (c) Maximum wind speed ($V_{max}$) and mean radius of 34-kt wind ($\overline{R}_{34kt}$). Identical analyses for (d-f) Ike (2008, AL); (g-i) Rita (2005, AL); (j-l) Michael (2018, AL); and (m-o) Sandy (2012, AL). Note the change in scale for $\overline{R}_{34kt}$ for Sandy in (o). Vertical gray line in each time series denotes time of CONUS landfall. Date labels on map correspond to 00 UTC.}\label{fig:casestudies}
\end{figure*}

Lastly, we illustrate our model applied to case studies of five impactful historical hurricanes: Patricia (2015), Ike (2008), Rita (2005), Michael (2018), and Sandy (2012) (Figure \ref{fig:casestudies}). The cases are presented roughly in order from least complex to most complex in terms of lifecycle evolution. Overall our model does well to predict $P_{min}$ across all five cases, but it is insightful to discuss both successes and notable biases across the cases.

Patricia (2015; Figure \ref{fig:casestudies}a-c) was a Category 5 storm in the eastern North Pacific that made landfall in Jalisco in western Mexico as a Category 4 hurricane (Figure \ref{fig:casestudies}a). Patricia was the most intense hurricane on record based on $V_{max}$ and exhibited the fastest rate of 24-hour intensification on record \citep{Kimberlain_Blake_Cangialosi_2016}. Hence it represents an excellent test case for extreme maximum wind speeds for our model. Our model does a remarkably good job of predicting the minimum pressure throughout its entire lifecycle (Figure \ref{fig:casestudies}b). The model exhibits near zero error during its initial period of intensification through 12 UTC 22 October as well as at its peak intensity (lowest pressure) at 12 UTC 23 October, with a modest high bias of approximately +5 hPa during the intervening 18 hours of very rapid intensification. In this case, the evolution of $P_{min}$ is driven almost entirely by $V_{max}$, as size ($\overline{R}_{34kt}$) increases only very gradually with time (Figure \ref{fig:casestudies}c). Additionally, Patricia remained at a relatively constant latitude, moving poleward by only 5$^o$ prior to landfall.

Ike (2008; Figure \ref{fig:casestudies}d-f) was a Category 4 storm in the North Atlantic that made multiple landfalls along the coasts of the Turks and Caicos Islands, Cuba, and finally Texas (Figure \ref{fig:casestudies}d). Ike grew substantially in size throughout its lifecycle while its intensity fluctuated (Figure \ref{fig:casestudies}f), and hence is a good test case of a storm that exhibited significant variations in both intensity and size. Our model does an excellent job of predicting the evolution of $P_{min}$ (Figure \ref{fig:casestudies}e). The model had a high bias (too weak) of approximately +10 hPa from 18 UTC 8 September to 06 UTC 11 September, a period during which Ike was passing along the long axis of Cuba. Note that many of those data occurred when the storm center was just offshore but close enough that the inner core of the storm almost certainly was partially onshore resulting in strong interaction with land \citep{Berg_2009}; the complexity of this interaction likely explains the discrepancy during that period. Thereafter, an eyewall replacement cycle occurred on 10 September that corresponded with a period of significant expansion through 12 UTC on 11 September while the maximum wind speed remained steady. At 00 UTC on 11 September the minimum pressure dropped and then increased very slowly to 950 hPa through 12 UTC, whereas our model predicted a more gradual decrease in pressure during this period. Finally, through landfall at 06 UTC on 13 September, the observed minimum pressure remained relatively constant, hovering within ~5 hPa of its final landfall pressure along Galveston Island, Texas of 950 hPa. Our model, on the other hand,  predicted a continued decrease in minimum pressure due to the slight intensification during 00-06 UTC on 12 September combined with its gradual poleward movement. Given the storm's very large size as it approached land, there is likely greater uncertainty in the estimate of the $\overline{R}_{34kt}$ in the model, which may explain the discrepancy during this period.

Rita (2005; Figure \ref{fig:casestudies}g-i) was a Category 5 storm in the North Atlantic that followed a very similar track as Ike through the northern Caribbean and Gulf of Mexico, except shifted slightly northward such that it passed through the Straits of Florida to the north of Cuba rather than directly over Cuba (Figure \ref{fig:casestudies}g). Rita also expanded steadily after interacting with Cuba. In contrast to Ike, during this expansion period Rita intensified rapidly, from 00 UTC on 21 September through 00 UTC on 22 September (Figure \ref{fig:casestudies}i). The final combination of extreme intensity and large size yielded the lowest Atlantic $P_{min}$ on record for the Gulf of Mexico (895 hPa). Our model performed very well over most of Rita's lifecycle (Figure \ref{fig:casestudies}h), including near-zero bias through the period of intensification and expansion and at peak intensity from 00-06 UTC 22 September. Thereafter, our model briefly exhibited a moderate high bias from 18 UTC 22 September through 00 UTC 23 September and then performed well with a smaller bias of 5-10 hPa leading up to landfall. The landfall pressure of 937 hPa was the lowest on record in the Atlantic basin for an intensity of 100 kt \citep{Knabb_Brown_Rhome_2006}, owing to its large size. Note that Rita's landfall intensity was slightly higher than Ike (100 kt vs. 95 kt), but Rita was also slightly smaller than Ike. Rita's landfall pressure was lower than Ike's landfall pressure, whereas our model predicted the opposite. Hence the model had a low bias for Ike and a high bias for Rita at landfall. Similar to Ike, Rita's large size as it approached land likely created larger uncertainty in the value of $\overline{R}_{34kt}$. Overall, Rita and Ike demonstrate how, for larger storms, size contributes significantly to reducing the minimum pressure relative to what would be expected from maximum wind speed alone. Rita and Ike made landfall at very similar latitudes (29.7$^o$N for Rita and 29.3$^o$N for Ike), so Coriolis played effectively no role in the differences between these two storms at landfall. As noted at the start of the Introduction, these large hurricanes with low $P_{min}$ relative to $V_{max}$ can cause confusion when considering how to communicate the severity of a storm, which is especially problematic when approaching land.

Michael (2018; Figure \ref{fig:casestudies}j-l) was a Category 5 storm in the North Atlantic that made landfall in the Florida Panhandle at its peak intensity of 140 kt. The system moved north throughout its lifecycle prior to landfall from its genesis location in the far western Caribbean (Figure \ref{fig:casestudies}j). Michael was similar to Patricia in that it reached extreme intensities while its size remained relatively constant, but it did so by intensifying more gradually and uniformly over a 4-day period leading up to landfall (Figure \ref{fig:casestudies}l). Our model did very well in predicting the continuous decrease in $P_{min}$ with time, particularly over its first few days prior to 00 UTC on 9 October during which the model had near zero error (Figure \ref{fig:casestudies}k). Our model exhibited a moderate low bias (too intense) of -10-15 hPa during the final two days leading up to landfall, with the error returning to near zero just prior to landfall. At 18 UTC on Oct 8, Michael passed very close to the western tip of Cuba and thereafter experienced a decay in its eyewall structure \citep{Beven_Berg_Hagen_2019}. This interaction with land may have induced uncertainties in wind radii estimates and changes in Michael's structure that were not captured by our model.

Finally, Sandy (2012; Figure \ref{fig:casestudies}m-o) was a highly destructive storm in the North Atlantic that became the largest storm ever recorded in the basin as measured by $\overline{R}_{34kt}$. Sandy made multiple landfalls in Jamaica, Cuba, and finally New Jersey as a post-tropical cyclone ~2.5 hours after completing extratropical transition (Figure \ref{fig:casestudies}m). Sandy represents perhaps the most complex of test cases, with dramatic changes in intensity, size, and latitude throughout its lifecycle (Figure \ref{fig:casestudies}o), multiple sustained interactions with land, and substantial extratropical interactions. Our model does well in predicting $P_{min}$ during its early stages in the southern Caribbean prior to crossing Cuba at 06 UTC on 25 October (Figure \ref{fig:casestudies}n). From 12 UTC on 25 October through 18 UTC on 26 October, our model exhibited a moderate low bias (too intense) of -10-20 hPa, particularly as the storm center crossed Cuba and then subsequently moved along the axis of the Turks and Caicos and the Bahamas. This period of time was associated with a sharp leveling off and then decrease in $V_{max}$ concurrent with a very rapid expansion in storm size. This expansion was driven by strong interaction with an upper-level trough and later a warm frontal boundary, which resulted in the development of an unusual structure with an exceptionally large $R_{max}$, and the strongest winds found on the western semicircle \citep{Blake_etal_2013}. The model performed better from 00-12 UTC on 27 October before the bias increased again to -15 hPa as Sandy moved back out over the open water while rapidly undergoing extratropical transition. Sandy was formally designated as an extratropical system on 2100 UTC on the 29th as the storm approached the New Jersey coastline, and indeed our model exhibited a very large low bias (far too intense) during 29 October, predicting a minimum pressure below 900 hPa when the actual value was 940 hPa prior to landfall. Hence, Sandy demonstrates how our model is not well-suited for storms that have undergone substantial extratropical transition. However, it is also worth noting that the estimation of the input parameter values ($V_{max}$ and $\overline{R}_{34kt}$) are likely highly uncertain during this latter period given that Sandy was a large and highly asymmetric storm whose wind field significantly overlapped with adjacent land.

Note that the biases in the above cases, particularly Ike and Michael, tend to be persistent for periods of 1-2 days when they occur. Such biases may be driven by a variety of factors. Estimating $R_{34kt}$ for larger storms may be more uncertain given that aircraft observations extend only 200 km from the center, so estimates depend more strongly on scatterometry data and can only be updated when new data become available. Land interaction may induce low biases in size as noted above. Additionally, there is uncertainty in estimating the environmental pressure $P_{env}$ from $P_{oci}$, as the closure of isobars depends on the synoptic-scale atmospheric flow on the periphery of the storm that tends to vary more slowly in time. Evaluating the relative role of uncertainties lies beyond the scope of this work. Future work might seek to develop more precise estimates of the environmental pressure, such as our method used above based on global analyses, that can be used both in an operational setting and in the Best Track archive.

\section{Conclusions}\label{sec:conclusions}

A simple model to predict $P_{min}$ in a tropical cyclone from routinely-estimated data would be useful for operations and practical applications. This work has developed an empirical linear model for the pressure deficit that takes as input the maximum wind speed, the mean radius of 34-kt wind, and storm central latitude, as well as the environmental pressure. The specific model predictors, given by $V_{max}^2$, $\frac{1}{2}fR_{34kt}$, and the ratio of the latter to the former, are derived from gradient wind balance applied to a modified Rankine vortex. Rather than using the full analytic solution itself, which will bake in the approximating assumptions of the theory, we instead use a simple linear regression model on those three parameters and estimate the coefficients from historical data: aircraft-estimated central pressure, historical Best Track data for storm wind field parameters, and global analyses for environmental pressure. This choice enables us to exploit the unique benefits of both physical theory (predictors) and of data (dependence on predictors). The model shows excellent skill, capturing 94.4\% of the variance in the historical dataset and outperforming various other versions of the model. It is superior to a standard wind-pressure relationship (89.4\% variance explained), which indicates that the inclusion of size and latitude information captures approximately half of the residual variance that is unexplained by this simpler model. Though the model was fit to data over the open ocean in the western tropical Atlantic in order to minimize the effects of observational uncertainties, it also appears to perform well at high latitudes and near coastlines. Finally, it performs very well when applied solely using data from the Extended Best Track database (94.9\% variance explained), which indicates it is viable in an operational setting using only routinely-estimated data.

The final model for the pressure deficit is given by Eq \eqref{eq:mlr_final}. The pressure deficit prediction is then translated to a prediction for $P_{min}$ via Eq. \eqref{eq:PmindP} by adding an estimate of the environmental pressure, which may be estimated operationally using the pressure of the outermost closed isobar via Eq \eqref{eq:Penv}.

Overall, this simple and fast model can predict $P_{min}$ from $V_{max}$, storm size, storm central latitude, and environmental pressure in practical applications.  In operations, the model could give a preliminary estimate of $P_{min}$ if other input predictors are available. Perhaps more importantly, $P_{min}$ can be calculated for up to 72 hours from the operational forecast time in the North Atlantic and eastern North Pacific given that the National Hurricane Center forecasts $V_{max}$,  and storm central position operationally out to 120 hours and 34-kt wind radii out to 72 hours. Currently, the National Hurricane Center does not operationally forecast $P_{min}$, but this simple tool would allow for a simple estimate of $P_{min}$ which would not entail additional work for the forecaster on duty. This information combined with radius of maximum wind estimates using similar methods \citep{Chavas_Knaff_2022,Avenas_etal_2023} could particularly be useful for storm surge model initial conditions. 

In risk analysis, the model can help provide mutually-consistent estimates of these three parameters that extend to other/future climate states where no observational data exist. In weather and climate modeling, the model could provide a new tool to evaluate the representation of TC intensity, size, and $P_{min}$ jointly, as well as to understand how future changes in $P_{min}$ reflect changes in intensity vs. size. More broadly, given the strong correlation between $P_{min}$ and historical economic damage, the model offers a full quantitative bridge from the wind field to hazards to damage that may be of use in the study of damage risk in both real-time forecasting and under climate change. Moreover, this model can help explain why larger storms can make it difficult to communicate the potential severity of a storm when the $P_{min}$ is substantially lower than expected for a given maximum wind speed (and Saffir-Simpson Hurricane Wind Sale category) based on a standard wind-pressure relationship.

Finally, we note that improved parameter estimation, particularly of $R_{34kt}$ and of the environmental pressure, may both help improve model performance for individual cases. As this method relies on operational estimates of $V_{max}$, $R_{34kt}$, and $P_{env}$, any improvements in those estimates will help in predicting $P_{min}$. Our estimates of $P_{env}$ can certainly be improved by more direct methods using global model analyses. Uncertainties associated with $V_{max}$ and $R34$ are on the order of 5 $ms^{-1}$ and 26 $km$, respectively, \citep{TornandSnyder2012,SampsonEA2017,Combot_etal_2020} will likely persist as few observational platforms exist that accurately estimate $R34$ and $V_{max}$ \citep{Knaff_etal_2021}. 




%

%

\clearpage
\acknowledgments

D. Chavas acknowledges funding support from NSF AGS grant 1945113. P. Klotzbach acknowledges funding support from the G. Unger Vetlesen Foundation. J. Knaff thanks his employer, NOAA Center for Satellite Applications and Research, for supporting this work. The scientific results and conclusions, as well as any views or opinions expressed herein, are those of the author(s) and do not necessarily reflect those of NOAA or the Department of Commerce.

%
%
\datastatement

All data used in this study will be made publicly available with DOI at the Purdue University Research Repository (PURR). The code and data associated with the initial submission have been posted at PURR at \url{http://data.datacite.org/10.4231/NKX4-MM81}.

%

\appendix

Here we derive the analytic solution for the pressure deficit, $\Delta P$, at the center of the storm that yields the three physical parameters used in the model presented in the main text. \citet{Klotzbach_etal_2022} showed that a modified-Rankine model well reproduces the relationship between $R_{max}$ and $R_{64kt}$, $R_{50kt}$, and particularly $\overline{R}_{34kt}$ in the EBTRK database over the tropical western Atlantic. This empirical outcome combined with the analytic simplicity of a modified Rankine wind profile motivate its use below.

The pressure deficit relative to a fixed radius can be expressed using gradient wind balance \citep{Knaff_Zehr_2007,Chavas_Reed_Knaff_2017}, given by:
\begin{equation}\label{eq:deltp}
    \Delta P = -\int_0^{R_o} \rho \left(\frac{v^2}{r} + fv\right) dr
\end{equation}
where $\rho$ is air density; $v$ is the rotating wind speed; $f=2\Omega sin(\phi)$ is the Coriolis parameter at the storm center with $\Omega = 7.292*10^{-5} s^{-1}$ and $\phi$ the latitude of the storm center; $r$ is radius ($r=0$ is the storm center); and $R_o$ is some larger radius. $\Delta P$ is defined as negative when the pressure decreases moving inwards, such that a stronger storm will have a more negative pressure deficit.

\begin{figure}[t]
 \centering
 \noindent\includegraphics[width=0.5\textwidth]{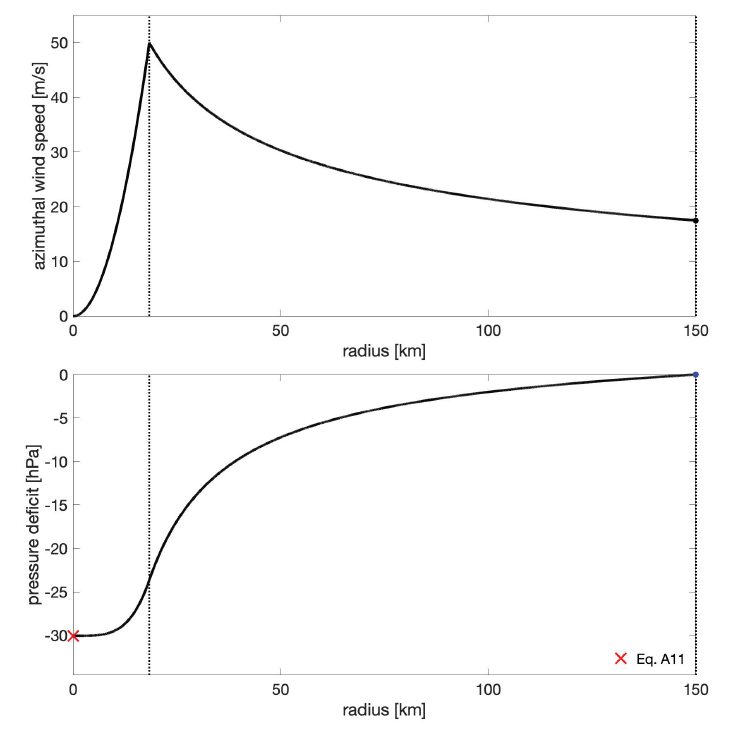}\\
 \caption{Example calculation of pressure deficit, $\Delta P$, inside of $\overline{R}_{34kt}$ for the wind profile used in this study based on a modified-Rankine structure, $v/V_{max} = \left(r/R_{max}\right)^\alpha$. (a) Wind profile, with $\alpha_e=2$ in the eye ($r \le R_{max}$) and $\alpha_o = -0.5$ outside of the eye ($r \ge R_{max}$); (b) Pressure deficit profile calculated from gradient wind balance and prediction from Eq. A11 (red x), with density assumed constant at $\rho = 1 \; kg/m^3$.}\label{fig:modrank}
\end{figure}

If given a profile of $v$ as a function of radius, Eq. \eqref{eq:deltp} can predict how pressure decreases moving inwards towards the center, including the total pressure deficit. A simple model for the wind profile inside of $\overline{R}_{34kt}$ makes use of the modified-Rankine equation:
\begin{equation}\label{eq:modrank}
    \frac{v}{V_{max}} = \left(\frac{r}{R_{max}}\right)^\alpha
\end{equation}
where $V_{max}$ is the maximum wind speed and $R_{max}$ is the radius of maximum wind. In the eye region ($r \le R_{max}$), we take $\alpha_e=2$, i.e. a parabola. Outside of the eye region ($r \ge R_{max}$), we take $\alpha_o = -0.5$, which is close to the best-fit value of -0.55 estimated from Extended Best Track data in the southwest Atlantic Ocean shown in Fig. 8a of \cite{Klotzbach_etal_2022}. An example of the pressure deficit profile calculated numerically from this wind profile using Eq \eqref{eq:deltp} at $\phi=20 \; ^o$N is shown in Figure \ref{fig:modrank}.

The simplicity of the wind model allows one to directly derive an analytic solution for $\Delta P$ by combining Eqs. \eqref{eq:deltp} and \eqref{eq:modrank}. To do so, it is helpful to first write Eq. \eqref{eq:deltp} in terms of $\tilde{v} = v/V_{max}$ and $\tilde{r} = r/R_{max}$, since the wind model (Eq. \eqref{eq:modrank}) is also written in this form, i.e.: 
\begin{equation}\label{eq:modrank_nondim}
    \tilde{v} = \tilde{r}^\alpha    
\end{equation}
We further take $\rho$ to be constant. Doing so yields:
\begin{equation}\label{eq:deltp_nondim}
    \Delta P = \rho V_{max}^2 \left(\int_0^{\tilde{R}_o} \frac{\tilde{v}^2}{\tilde{r}}d\tilde{r} + f\int_0^{\tilde{R}_o} \tilde{v} d\tilde{r}\right)
\end{equation}
Using Eq. \eqref{eq:modrank_nondim} and Eq. \eqref{eq:deltp_nondim}, $\Delta P$ can be calculated analytically both within the eye region ($\Delta P_e$; $\tilde{r}<1$) and outside the eye region ($\Delta P_o$; $\tilde{r}>1$), with the total pressure deficit equal to the sum of the two:
\begin{equation}\label{deltp_tworegion}
    \Delta P = \Delta P_e + \Delta P_o
\end{equation}

The general solution for any $\alpha$ is given by
\small
\begin{equation}
    \Delta P_e = -\left(\frac{\rho}{2\alpha_{e}}\right)V_{max}^2 - \left(\frac{2\rho}{(\alpha_e+1)V_o^{\frac{1}{\alpha_o}}}\right)\left(\frac{1}{2}fR_o\right)V_{max}^{1+\frac{1}{\alpha_o}}
\end{equation}
\normalsize
within the eye, and
\tiny
\begin{equation}
    \Delta P_o = -\left(\frac{\rho}{2\alpha_o}\right)\left(V_o^2 - V_{max}^2\right) - \left(\frac{2\rho}{(\alpha_e+1)V_o^{\frac{1}{\alpha_o}}}\right)\left(\frac{1}{2}fR_o\right)\left(V_o^{1+\frac{1}{\alpha_o}}-V_{max}^{1+\frac{1}{\alpha_o}}\right)
\end{equation}
\normalsize
within the outer region. We plug in $\alpha_e = 2$ and take $\alpha_o = -0.5$, which is close to the best estimate of -0.55 and makes the math for the final solution much simpler. Doing so yields
\begin{equation}
    \Delta P_e = -\frac{1}{4}\rho V_{max}^2 - \frac{2}{3}\rho V_o^2\left(\frac{\frac{1}{2}fR_o}{V_{max}}\right)
\end{equation}
within the eye, and
\small
\begin{equation}
    \Delta P_o = -\rho V_{max}^2 + \rho V_o^2 - 4V_o\left(\frac{1}{2}fR_o\right) + 4V_o^2\left(\frac{\frac{1}{2}fR_o}{V_{max}}\right)
\end{equation}
\normalsize
within the outer region. Adding the two together and combining like terms yields a final equation for the total pressure deficit:
\tiny
\begin{equation}\label{eq:dp_analytic}
    \Delta P_{theo} = \rho \left[V_0^2 -\frac{5}{4}\left(V_{max}^2\right) - 4V_o\left(\frac{1}{2}fR_o\right) + \frac{10}{3}V_o^2\left(\frac{\frac{1}{2}fR_o}{V_{max}}\right) \right]
\end{equation}
\normalsize
This equation is a linear function of three predictors: $V_{max}^2$, $\frac{1}{2}fR_o$, and $\frac{\frac{1}{2}fR_o}{V_{max}}$. The first term ($\rho V_0^2$) is a constant, as are the other quantities in Eq. \eqref{eq:dp_analytic}. These three predictors are composed of parameters that are routinely estimated operationally. Note that the model predicts specifically that $\Delta P$ is larger in magnitude (more negative) for larger $V_{max}^2$ and $\frac{1}{2}fR_o$, but it is smaller in magnitude for larger $\frac{\frac{1}{2}fR_o}{V_{max}}$. Plugging in $\rho=1 \; kg/m^3$ and $V_o = 17.5 \; ms^{-1}$ and dividing through by 100 to convert to hPa yields:
\tiny
\begin{equation}\label{eq:dp_analytic_numerical}
    \Delta P_{hPa,theo} = 3.0625 -0.0125\left(V_{max}^2\right) - 0.70\left(\frac{1}{2}fR_o\right) + 10.208\left(\frac{\frac{1}{2}fR_o}{V_{max}}\right)
\end{equation}
\normalsize

Note that the final term ($\frac{\frac{1}{2}fR_o}{V_{max}}$) offsets the contribution to the pressure deficit from the first two terms: the cyclostrophic pressure drop from the wind profile (first term) and the Coriolis pressure drop from the overall storm wind profile out to $R_o$ (second term). This final term represents a small additional pressure drop contribution from the Coriolis term arises because $R_{max}$ contracts inwards with intensification (i.e. this term becomes less negative as $V_{max}$ increases and $R_{max}$ decreases), and thus the outer region expands inwards. Physically, this final term originates from the Coriolis term in the gradient wind balance equation -- it represents an inverse vortex Rossby number in front of this term, where $Ro = V_{max}/(fR_{max})$ is the Rossby number at $R_{max}$ but where the modified Rankine vortex allows for $R_{max}$ to be substituted with $R_o$.

Why use the empirical model of Eq. \eqref{eq:mlr_basic} rather than directly using Eq. \eqref{eq:dp_analytic_numerical}? Three reasons:
\begin{enumerate}
    \item The value $\alpha_o = -0.5$ approximates the best-fit value of -0.55 to simplify the math.
    \item The derivation neglects the small pressure drop outside of $\overline{R}_{34kt}$.
    \item The derivation assumed constant density in order to be analytically tractible. That is not true though since P decreases with radius while temperature remains relatively constant \citep{Emanuel_1986}, and hence density decreases moving radially inwards following the Ideal Gas Law.
    \item The model assumes winds are taken at the boundary layer top, which is very difficult to estimate or even agree on a definition in practice. Real data for intensity and size are estimates near the surface. A wind speed reduction due to friction within the boundary layer will reduce wind speeds and hence will low-bias the pressure deficit, but accounting for this is very complex.
\end{enumerate}
Given these assumptions, it is preferable to use the theory solely to identify the most important physical parameters while allowing the empirical model to determine the dependencies (i.e. regression coefficients) found in nature, similar to the approach of \citet{Chavas_Knaff_2022} and \citet{Avenas_etal_2023} for the radius of maximum wind. Nonetheless, it is still conceptually instructive to compare the theoretical coefficients of Eq. \eqref{eq:dp_analytic_numerical} against the empirical values (Eq. \eqref{eq:mlr_final}) to ensure that they are at least consistent in their order of magnitude, which indeed they are. The lone notable minor exception is the value of the constant term, which is slightly positive in the theoretical model but slightly negative in the empirical model, which suggests a deeper pressure drop than predicted by theory. One simple explanation is that this difference of approximately 6 hPa represents the small pressure drop between $\overline{R}_{34kt}$ and the outer edge of the storm that is neglected in the theory above.





%



\bibliographystyle{ametsocV6}
\bibliography{refs_CHAVAS}

\end{document}